Origin of the anomalous magnetic circular dichroism spectral shape in ferromagnetic Ga$_{1-x}$Mn$_x$As: Impurity bands inside the band gap

K. Ando*, H. Saito, K. C. Agarwal, M. C. Debnath, and V. Zayets

National Institute of Advanced Industrial Science and Technology (AIST),

Nanoelectronics Research Institute,

Tsukuba Central 2, Umezono 1-1-1, Tsukuba, Ibaraki 305-8568, Japan

* e-mail: ando-koji@aist.go.jp

Abstract

The electronic structure of a prototype dilute magnetic semiconductor (DMS), Ga$_{1-x}$Mn$_x$As, is studied by magnetic circular dichroism (MCD) spectroscopy. We prove that the optical transitions originated from impurity bands cause the strong positive MCD background. The MCD signal due to the $E_0$ transition from the valence band to the conduction band is negative indicating that the *p-d* exchange interactions between the *p*-carriers and *d*-spin is antiferromagnetic. The negative $E_0$ MCD signal also indicates that the hole-doping of the valence band is not so large as previously assumed. The impurity bands seem to play important roles for the ferromagnetism of Ga$_{1-x}$Mn$_x$As.

The discovery of ferromagnetic order in the dilute magnetic semiconductor (DMS) $Ga_{1-x}Mn_xAs$ has generated a great deal of interest[1,2]. It is generally accepted that the holes donated by the substitutional Mn mediate the ferromagnetic order of the localized Mn spins through the exchange interaction between the carrier and the spin[2]. Despite a considerable number of experimental and theoretical studies since the discovery of the ferromagnetic $Ga_{1-x}Mn_xAs$, however, the electronic structure of this material, the character of the holes, and the mechanism of the ferromagnetic order in it are still not well understood[3–7].

Magneto-optical spectroscopy is a powerful tool for probing the electronic structures of DMS's[8-10]. Indeed, an unambiguous proof of the intrinsic nature of the ferromagnetism of $Ga_{1-x}Mn_xAs$ was provided by magnetic circular dichroism (MCD) spectroscopy[11]. Subsequently, many investigators have used magneto-optical spectroscopy in efforts to clarify the electronic structure of $Ga_{1-x}Mn_xAs$ in more detail[12–18].

According to the established MCD analysis[8] for II-VI DMS's, negative and positive MCD polarities at the $E_0$ critical point (CP) in the center of the Brillouin zone (the $\Gamma$ point) respectively correspond to antiferromagnetic and ferromagnetic *p-d* exchange interactions between the *p*-carriers and *d*-spin. The polarity of the MCD signal at $E_0+\Delta_0$ should be opposite that at $E_0$[8]. One therefore expects a pair of positive and negative MCD signals around the fundamental absorption edge. Beschoten *et al.*[12], however, reported that the MCD signal around the fundamental absorption edge of $Ga_{1-x}Mn_xAs$ was very broad and only positive. Since such a broad and unipolar signal is not seen in the MCD spectra of traditional insulating paramagnetic II-VI DMS's[8], its shape might provide key information about the electronic structure of $Ga_{1-x}Mn_xAs$.

The *p-d* exchange interaction of $Ga_{1-x}Mn_xAs$ is widely believed to be antiferromagnetic[19, 20]. To explain both the expected antiferromagnetic *p-d* exchange interaction and the observed positive MCD at $E_0$ consistently, Szczytko *et al.*[13] argued that the expected MCD polarity is inverted because the allowed optical transition is shifted away from the center of the Brillouin zone because the valence band (VB) contains many holes (the Moss-Burstein shift). According to this explanation, the Fermi level ($E_F$) is deep in the VB (the VB-hole model). However, this explanation was only qualitative [13,14]. The quantitative analyses of the observed shape of the MCD spectrum based on the VB-hole model forced Lang *et al.*[16] to conclude that the *p-d* exchange interaction was ferromagnetic which was opposite the expected one.

Ando *et al.*[11] and Beschoten *et al.*[12], on the other hand, although their interpretations of the MCD structures were not identical, considered the anomalous shape of the spectrum to be due to a faint negative MCD peak on a strong positive MCD background. But this explanation was not persuasive because the origin of the positive MCD background was not identified.

In this article we provide a clear explanation of the anomalous shape of the $Ga_{1-x}Mn_xAs$ MCD spectrum and its implications for electronic structures. Up to now, the reported MCD spectra of $Ga_{1-x}Mn_xAs$ were limited only to the photon energy range around the fundamental absorption edge. In the work reported here we measured the MCD spectra in a wider range, from 0.6 to 4 eV.

In order to measure the MCD spectra in the transmission configuration up to 4 eV and to avoid the optical interference effect, we used thin (50 nm) $Ga_{1-x}Mn_xAs$ films

[ 3 ]

grown by the molecular beam epitaxy method. The films were grown at 220–250°C on 20-nm-thick GaAs buffer layers grown at 450°C on sapphire (0001) substrates. No trace of a second phase was detected in the x-ray diffraction patterns of these (111)-oriented films. The Mn concentration $x$ was measured by an electron probe microanalysis, and the ferromagnetic Curie temperature $Tc$ determined by a superconducting quantum interference device (SQUID) was 35 K for x=0.030 and 20 K for x=0.022.

MCD signal was measured by detecting the difference between the absorptions of left- and right-handed circularly polarized light with the polarization modulation method[8]. A magnetic field was applied along the film normal, and the samples were cooled by a closed-cycle refrigerator.

Figure 1 shows the MCD spectrum of paramagnetic $Ga_{1-x}Mn_xAs$ (x=0.004). The MCD signal from the GaAs buffer layer was confirmed to be negligibly small. The spectrum is composed of four parts: structures in the range between 2.2 and 3.4 eV, a positive signal around 1.8 eV, a negative signal around 1.5 eV, and a structure below 1.4 eV. The reported energies of the $E_0$, $E_0+\Delta_0$, $E_1$ and $E_1+\Delta_1$ CP's of GaAs[21] are shown in Fig. 1 by dashed lines.

MCD is generally prominently enhanced around the CP's of the band structure[8]. Correspondences with the CP energies show that the MCD structures between 2.2 and 3.4 eV— a positive peak around 3 eV and a negative wing from 2.2 to 2.8 eV—are due to the optical transitions at the L point of the Brillouin zone.

The inset in Fig. 1 shows the MCD spectrum of insulating paramagnetic $Cd_{1-x}Mn_xTe$ (x=0.08) as a reference[8]. The horizontal scales are adjusted so that the positions of the $E_0$ and $E_0+\Delta_0$ CP's of GaAs and $Cd_{1-x}Mn_xTe$ coincide. The shape of the MCD spectrum of $Ga_{1-x}Mn_xAs$ is very similar to that of the MCD spectrum of $Cd_{1-x}Mn_xTe$. This correspondence shows clearly that the MCD structures of $Ga_{1-x}Mn_xAs$ (x = 0.004) at around 1.5 eV and 1.8 eV are respectively due to the $E_0$ and $E_0+\Delta_0$ transitions.

The MCD structure below 1.4 eV has not been reported before. Since its energy is below the band-gap energy of GaAs, it must be produced by transitions related to the impurity levels inside the band gap. One possible impurity level is the arsenic antisite ($As_{Ga}$). The MCD signal of bulk GaAs around 0.9 eV has been attributed to the photoionization from the $As_{Ga}$ level to the VB[22]. Another possible impurity level is the Mn(0/+) level of the interstitial Mn. Although this level has not yet been observed experimentally, a theoretical study[23] has predicted it to be 0.98 eV above the top of the VB. A transition from the Mn acceptor level located about 0.11 eV above the top of the VB[24] to the conduction band might also contribute to the MCD signal. Further study is needed to clarify the fine structures of the impurity-level-related contributions to MCD.

Figure 2 shows the MCD spectra of $Ga_{1-x}Mn_xAs$ samples with various Mn contents. The MCD spectrum reported by Beschtoen et al. [12] of a sample with $x$ = 0.053, $T_c$ = 101 K, and a hole concentration of 1.5 x $10^{20}$ $cm^{-3}$ is also shown by a dotted-curve. Its sharp MCD structures are due to the better crystalline quality of the sample grown on the lattice-matched GaAs substrate. The samples of $x$ = 0.022 and $x$ = 0.030 showed the ferromagnetic magnetic field dependence as expected from the SQUID data. The MCD spectra of each sample measured at different magnetic field and a fixed temperature were confirmed to collapse onto the universal shape when normalized by the MCD intensity. This implies that the observed MCD signal comes from one magnetic material, and is free from the contribution of possible second phase materials[9, 25].

The MCD signal of the high-Mn content samples is positive in the range between 1.4 and 2.4 eV as reported in Ref. 12. But it should be noted that the characteristic features seen in the MCD spectrum of a sample with a low Mn content (Fig. 1)—*i.e.*, a minimum around 1.5 eV and a maximum around 1.8 eV—are seen in the MCD spectra of all the samples.

The shape of the MCD spectrum below the absorption edge changes with Mn content, indicating that the plural impurity bands contribute the MCD signal. The contribution from the structure around 1 eV rapidly increased with increasing Mn content and eventually became a very broad peak with a long tail on the higher-energy side. Optical transitions between the VB and impurity bands (IB's) in GaAs are known to have long tails on the higher-energy side because of the relaxation of the optical selection rule[7, 26].

It became clear that the anomalous positive MCD signal[12] around the absorption edge of $Ga_{1-x}Mn_xAs$ is composed of a negative peak at $E_0$, a positive structure around $E_0+\Delta_0$, and a very broad positive background due to the IB-related transitions (Fig. 3). A negative contribution from the *L* point also contributes to the signal in the range above 2.2 eV (Figs. 1 and 2). Beschoten *et al.*[12] argued that a faint negative MCD peak was hidden in the MCD shoulder at 1.6 eV rather than at 1.5 eV. This misinterpretation was due to the positive peak around 1.8 eV being mistakenly assigned to the Mn intraionic transition[12]. Its actual origin is the $E_0+\Delta_0$ transition, and the Mn intraionic transitions in DMS's should not cause a large MCD signal[27]. Subtle difference of the temperature dependences of the MCD contributions form the $E_0$ and $E_0+\Delta_0$ transitions may have caused the spurious negative MCD peak in their analysis.

The negative MCD at the $E_0$ of $Ga_{1-x}Mn_xAs$ clearly shows the antiferromagnetic nature of the *p-d* exchange interaction[8]. Further detailed studies with thicker samples will enable us to evaluate the strength of this interaction quantitatively.

As indicated by Szczytko *et al.*[13], the polarity and the position of the MCD structures should be closely related to the occupancy of the VB by electrons. Our analyses show that both the polarity and shape of the MCD spectrum due to the interband transition at the Γ point ($E_0$ and $E_0+\Delta_0$) of ferromagnetic $Ga_{1-x}Mn_xAs$ are essentially same as those of the MCD spectrum due to the corresponding transitions in insulating paramagnetic $Cd_{1-x}Mn_xTe$ (Figs.1–3). This indicates that the hole doping of VB is too small to modify the shape of the MCD signal due to the $E_0$ CP. As shown in Fig.4, the position of the MCD minimum around $E_0$ does not show noticeable change with Mn-content. Slightly lower energies observed for $x = 0.022$ and $x = 0.030$ may be due to the broadening of the MCD structures. The energy level and the electron occupancy of the VB of $Ga_{1-x}Mn_xAs$ seem to be rarely affected by the Mn substitution. This is in a sharp contrast with a widely used phenomenological model of $Ga_{1-x}Mn_xAs$ proposed by Dietl *et al.*[3], who assumed the Fermi level locates below the top of the valence band (VB-hole model). Theoretical analyses[14, 28] based on the VB-hole model predicted that the energy positions of the MCD structures should shift to the higher energy side with increasing Mn content because a greater Mn content would make the $E_F$ deeper in the VB. However, such blue-shifts were not observed experimentally (Fig. 4).

The essential difference between the MCD spectra of ferromagnetic $Ga_{1-x}Mn_xAs$ and paramagnetic $Cd_{1-x}Mn_xTe$ is the broad MCD signal due to the IB's in $Ga_{1-x}Mn_xAs$ (Figs. 1-3). Up to now, only three ferromagnetic DMS's have been confirmed to be intrinsic[9]: $Ga_{1-x}Mn_xAs$[11], $In_{1-x}Mn_xAs$[29], and $Zn_{1-x}Cr_xTe$[30]. It is interesting that $Zn_{1-x}Cr_xTe$, which has a *Tc* of 300 K, also shows a broad MCD signal below its fundamental absorption edge[29, 31]. IB's may generally play important roles for the appearance of ferromagnetic order in DMS's.

The carrier in the IB is another possible candidate[7,32] which is responsible the ferromagnetism of $Ga_{1-x}Mn_xAs$. Our experimental results seem to be consistent with the IB-hole model, which seems to be also supported by the angle-resolved photoemission spectroscopy[33], the resonant tunneling spectroscopy of quantum-wells[34], and the infrared ellipsometry[7,35]. Further MCD spectroscopy studies of the IB's in DMS's will enable us to clarify the electronic structure of $Ga_{1-x}Mn_xAs$ and to find a guiding principle to make high-performance ferromagnetic DMS's.

In conclusion, we have shown that the anomalous shape of the MCD spectrum of ferromagnetic dilute magnetic semiconductor $Ga_{1-x}Mn_xAs$ is due to a newly discovered broad and positive MCD background signal caused by the optical transitions related to impurity bands inside the band gap. Our analyses of MCD spectra revealed an antiferromagnetic *p-d* exchange interaction in $Ga_{1-x}Mn_xAs$ and suggests that the carriers in the impurity bands play important roles for the ferromagnetism of this material.

Acknowledgements
We thank Dr. A. K. Bhattacharjee for useful discussions and Ms. A. Yamamoto for contributions to the sample preparation.

Figure captions

Fig. 1.   MCD spectrum of paramagnetic $Ga_{1-x}Mn_xAs$ (x=0.004) at 6 K and 1 T. Reported energies[21] of the critical points of GaAs are shown by broken vertical lines. The inset shows the MCD spectrum of paramagnetic $Cd_{1-x}Mn_xTe$ (x=0.08)[8] at 15 K and 1 T. The horizontal scales are adjusted so that the $E_0$ and $E_0+\Delta_0$ positions of the two materials coincide.

Fig. 2.   MCD spectra, at 6 K and 1 T, of $Ga_{1-x}Mn_xAs$ samples with various Mn contents. The spectra of paramagnetic samples (x = 0.004 and 0.011) are magnified 4-fold for clarity. A MCD spectrum reported by Ref. 12 is shown by a red dotted curve in an arbitrary unit vertical scale.

Fig. 3.   Schematic decomposition of the anomalous positive MCD signal around the absorption edge of $Ga_{1-x}Mn_xAs$. The negative $E_0$ peak, positive $E_0+\Delta_0$ peak, and negative L signal are due to transitions between the valence band and conduction band. The broad and positive MCD signal is due to transitions to impurity bands (IB's).

Fig. 4.   Mn content dependence of the photon energies corresponding to the minimum and maximum of the MCD intensity at 6 K. The data from Ref. 12 are also shown by open circle and box. Reported energies[21] of the critical points of GaAs are shown by broken horizontal lines.

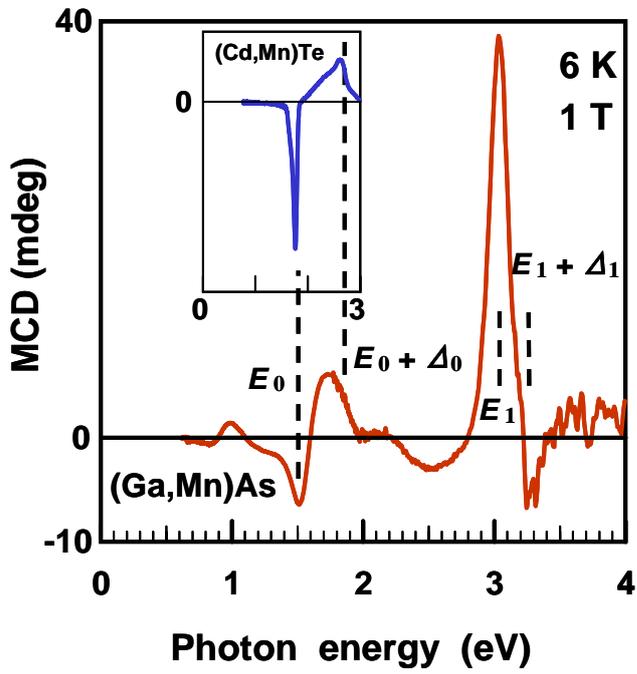

Fig.1

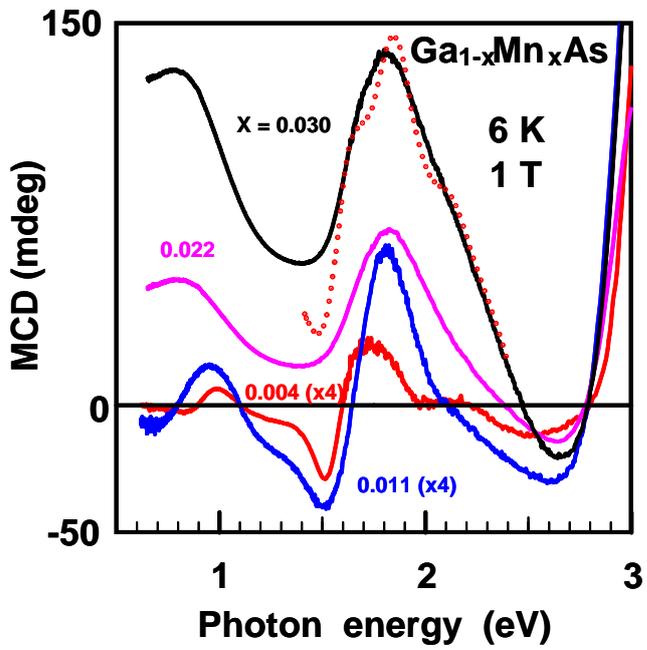

Fig.2

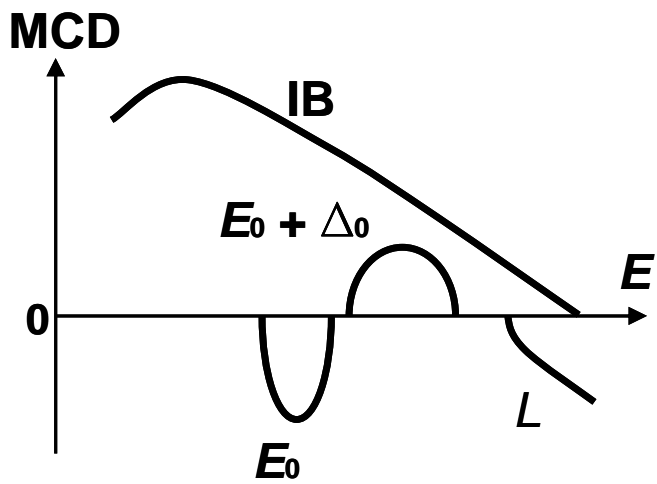

Fig.3

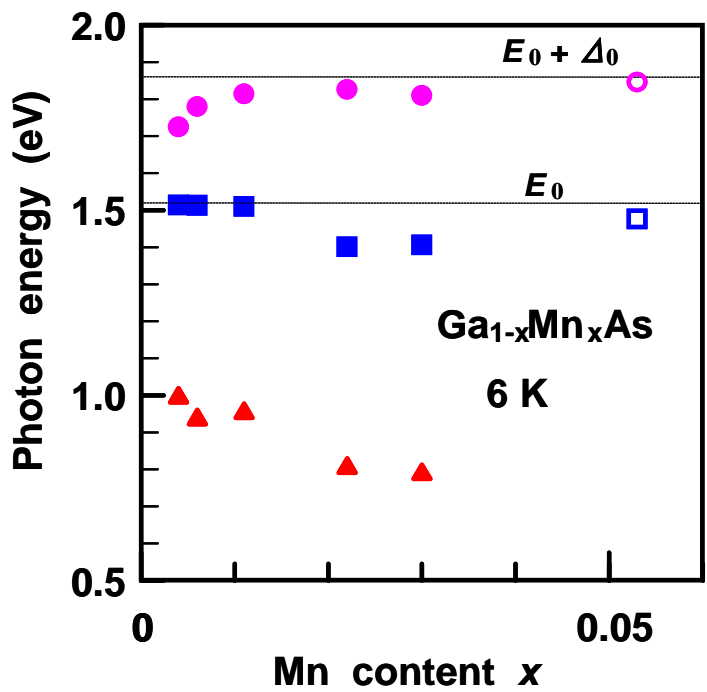

Fig.4